\documentclass[journal=jacsat,manuscript=article]{achemso}

\usepackage{chemformula} 
\usepackage[T1]{fontenc} 

\usepackage{color}
\newcommand{\red}[1]{{\color{black} #1}}

\author{Titus Masese}
\affiliation{Research Institute of Electrochemical Energy, National Institute of Advanced Industrial Science and Technology (AIST), 1-8-31 Midorigaoka, Ikeda, Osaka 563-8577, JAPAN}
\alsoaffiliation{AIST-Kyoto University Chemical Energy Materials Open Innovation Laboratory (ChEM-OIL), Sakyo-ku, Kyoto 606-8501, JAPAN}
\email{titus.masese@aist.go.jp}

\author{Yoshinobu Miyazaki}
\affiliation{Tsukuba Laboratory, Technical Solution Headquarters, Sumika Chemical Analysis Service (SCAS), Ltd.,Tsukuba, Ibaraki 300-3266, JAPAN}
\email{yoshinobu.miyazaki@scas.co.jp}

\author{Godwill Mbiti Kanyolo}
\affiliation{Department of Engineering Science, The University of Electro-Communications, 1-5-1 Chofugaoka, Chofu, Tokyo 182-8585, JAPAN}
\email{gmkanyolo@mail.uec.jp}

\author{Teruo Takahashi}
\affiliation{Tsukuba Laboratory, Technical Solution Headquarters, Sumika Chemical Analysis Service (SCAS), Ltd.,Tsukuba, Ibaraki 300-3266, JAPAN}

\author{Miyu Ito}
\affiliation{Tsukuba Laboratory, Technical Solution Headquarters, Sumika Chemical Analysis Service (SCAS), Ltd.,Tsukuba, Ibaraki 300-3266, JAPAN}

\author{Hiroshi Senoh}
\affiliation{Research Institute of Electrochemical Energy, National Institute of Advanced Industrial Science and Technology (AIST), 1-8-31 Midorigaoka, Ikeda, Osaka 563-8577, JAPAN}
\email{h.senoh@aist.go.jp}

\author{Tomohiro Saito}
\affiliation{Tsukuba Laboratory, Technical Solution Headquarters, Sumika Chemical Analysis Service (SCAS), Ltd.,Tsukuba, Ibaraki 300-3266, JAPAN}
\email{tomohiro.saito@scas.co.jp}

\title[An \textsf{achemso} demo]
  {\red{Topological Defects and Unique Stacking Disorders in Honeycomb Layered Oxide $\rm K_2Ni_2TeO_6$ Nanomaterials: Implications for Rechargeable Batteries}}

\begin{document}



\newpage


\newpage

\begin{figure*}[!ht]
\centering
  \includegraphics[width=13cm]{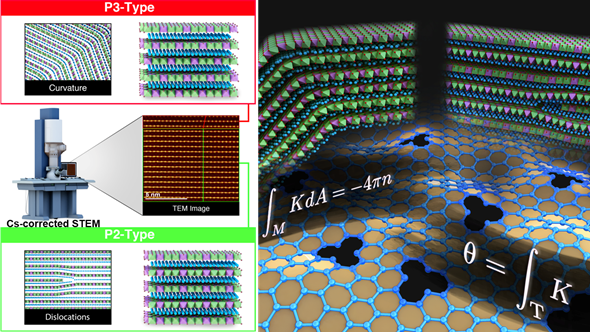}
\end{figure*}

\begin{abstract}
Endowed with a multitude of exquisite properties such as rich electrochemistry, superb topology and eccentric electromagnetic phenomena, honeycomb layered oxides have risen to the top echelons of science with applications in diverse fields ranging from condensed matter physics, solid-state chemistry, materials science, solid-state ionics to electrochemistry. However, these oxides are vastly underutilised as their underlying atomistic mechanisms remain unexplored. Therefore, in this study, atomic-resolution imaging on pristine $\rm K_2Ni_2TeO_6$ along multiple zone axes was conducted using spherical aberration-corrected scanning transmission electron microscopy (Cs-corrected STEM) to reveal hitherto unreported \red {nanoscale} topological defects and curvature which can be associated with various phase transitions. Furthermore, we discover the coexistence of \red {a stacking} variant with P3-type sequence alongside the well-reported P2-type stacking sequence in such honeycomb layered oxides. Our findings have the potential to inspire further experimental and theoretical studies into the role of stacking and topology in the functionality of honeycomb layered oxides \red {, for instance, as high-performance electrode materials for rechargeable batteries}.

\noindent{\red{Keywords: Honeycomb layered oxides, transmission electron microscopy (TEM), aberration-corrected STEM, topological defects, stacking disorders, dislocations, atomic-resolution imaging, curvature}}
\end{abstract}


\newpage

\section{Introduction}

In the quest for energy independence and ecological sustainability, honeycomb layered oxides\cite{Kanyolo2020a} have drawn enormous interest for their potential as rechargeable battery components owing to their fascinating two-dimensional (2D) ionic diffusion governed by phase transitions.\cite{Kanyolo2020a, Kanyolo2020b} These compounds generally adopt the chemical composition $A_2M_2D \rm O_6$, $A_3M_2D \rm O_6$ or $A_4MD \rm O_6$ (where $M$ can be divalent or trivalent transition metal atoms such as $\rm Cr, Mn, Fe$ $\rm Co, Ni, Cu$ or some combination thereof; $D$ represents pentavalent or hexavalent pnictogen or chalcogen metal atoms such as $\rm Te, Sb, Bi$; and $A$ can be alkali atoms such as $\rm Li, Na, K,$ \textit{etc} or coinage-metal atoms such as $\rm Cu, Ag,$ \textit{etc}.)\cite{Sathiya2013, Grundish2019, Yang2017, Yuan2014, Masese2018, Bhange2017, Gyabeng2017, Yadav2019, Yoshii2019, Zheng2016, Ma2015, Wang2019JMaterChemA, Dai2017, Evstigneeva2011, Kanyolo2020a, Kumar2012, Bhardwaj2014, Kumar2013, Berthelot2012a, Berthelot2012InorgChem, Berthelot2012JSolidStateChem, Berthelot2012, Viciu2007, He2017, He2018, Schmidt2013, Masese2019} Such structures comprise an array of transition metal sheets consisting of $D \rm O_6$ octahedra surrounded by multiple $M \rm O_6$ octahedra in a distinct hexagonal (honeycomb) alignment. 
\red {O}xygen atoms from the octahedra in turn coordinate with $A^+$ cations interposed between the layers to form heterostructures whose interlayer bonds are significantly weaker than the covalent in-plane bonds. 

Progress towards their battery application has been hindered by the scarcity of evidence for their unique topologies, \red {nanoscale} defects and curvature effects that ought to accompany their otherwise well-reported stacking sequences. Given that the interlayer distance is inversely proportional to the strength of the interlayer bonds, the size of the resident $A$ atoms (Shannon-Prewitt radii)\cite{Shannon1976} typically determines the structural topology and the resulting physicochemical properties of the honeycomb layered oxides. For instance, smaller atomic radii cations such as $\rm Li$ atoms in $\rm Li_2Ni_2TeO_6$ tend to form stronger tetrahedral coordinations between $\rm Li$ atoms and oxygen atoms, with 2 repetitive honeycomb layers in each unit cell. \cite{Grundish2019} This type of structure is typically referred to as T2-type in the Hagenmuller notation\cite{Delmas1976} (where `T' is the tetrahedral coordination and `2' is the number of honeycomb slabs per unit cell). \red{The research findings for other honeycomb layered oxide materials such as \red { $\rm H_3LiIr_2O_6$, $\rm Ag_3LiIr_2O_6$, $\rm Ag_3LiRu_2O_6$}, amongst others\cite{bette2017, bette2019, kitagawa2018, kimber2010, todorova2011} are consistent with a dumbbell coordination of alkali atoms with oxygen and are classified as D-type structures.}\cite{Kanyolo2020a} On the other hand, larger ionic radii atoms such as $\rm Na^+$ in $\rm Na_3Ni_2SbO_6$ adopt an O3-type layered framework (where `O' is an octahedral coordination with a periodicity of three honeycomb slabs per unit cell) whilst $\rm Na^+$ and $\rm K^+$ in $\rm Na_2Ni_2TeO_6$ and $\rm K_2Ni_2TeO_6$, respectively, adopt a P2-type framework (where `P' is a prismatic coordination with a periodicity of two). \cite{Yuan2014, Masese2018, Ma2015, Evstigneeva2011, Kanyolo2020a, matsubara2020sci} Amongst the assorted honeycomb layered oxides reported so far,\cite{Kanyolo2020a} P-type honeycomb layered oxides have been noted to encompass the widest interlayer distances. \cite{Sun2019} Thus, alkali atoms with larger ionic radii such as $\rm K$ or $\rm Rb$ typically result in P-type configurations. 

Amongst honeycomb layered oxides entailing chalcogen or pnictogen atoms, P3-type structures have been found to have manifold electrochemical capabilities associated with multiple topological defects resulting from their weak interlayer bonds. \cite{Wang2018, Kanyolo2020a, Dai2017, Ma2015} However, these structures are difficult to synthesise and analyse in as-prepared honeycomb layered oxide materials due to their inferior stability. Thus, theoretical and experimental explorations into the relationship between topological defects innate to these structures and emergent nanoscale features remain underdeveloped. Although X-ray diffraction (XRD) has widely been employed to provide the average structural information of crystalline materials, it cannot be used on pristine materials to account for short-range structural evolutions occurring immediately after synthesis. Consequently, transmission electron microscopy (TEM) can be employed alongside XRD analyses to provide local atomistic information of the cathode, for instance the intermittent stacking evolutions before and after electrochemical operations. The high precision of TEM complements XRD by
\begin{table}
\centering
\caption*{\textbf{Scheme 1: The solid state synthesis protocol of honeycomb layered oxide ($\rm K_2Ni_2TeO_6$) showing the various stages (See experimental details furnished as Supporting Information)}}
\includegraphics[width=\textwidth]{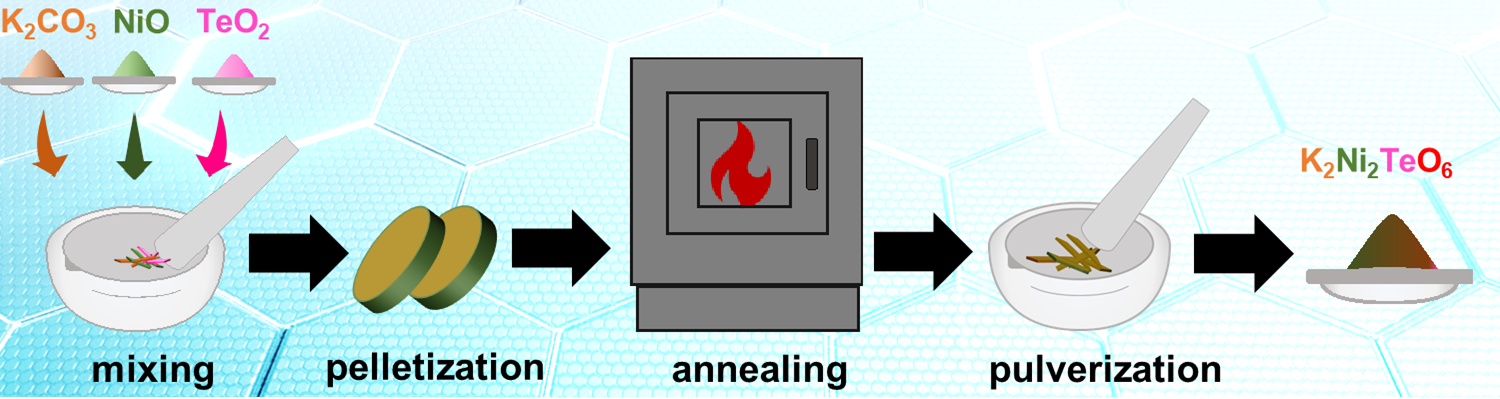}
\end{table}
identifying defects (disorders) beyond the detectable limits of the latter.  
 
In order to ascertain the structural nature and topological behaviour of the enigmatic P-type honeycomb structures, we employ atomic-resolution scanning transmission electron microscopy (STEM) on pristine $\rm K_2Ni_2TeO_6$ cathode material \red {for rechargeable potassium-ion battery} (prepared using the protocol details as displayed in \red {Scheme 1}) to reveal emergent structural disorders along $c$-axis. Our analyses reveal partially missing honeycomb slabs of $\rm NiO_6$ and $\rm TeO_6$ octahedra as well as absence of some constituent $\rm K$ atoms. Furthermore, we identify regions comprising both P2- and P3-type stacking sequences, which correlate with unique topological defects and \red {nanoscale} curvatures triggering phase transitions during battery operations.

\red{\section{Results and Discussion}}

Polycrystalline samples of $\rm K_2Ni_2TeO_6$ were synthesised through the conventional high-temperature solid-state ceramics route (details provided in the \red {EXPERIMENTAL} section). The composition of the as-prepared material was ascertained using inductively coupled plasma-atomic emission spectroscopy (\red {Table S}1), whereas the uniformity of the constituent elements and morphology were validated using scanning electron microscopy (shown in \red {Figure S}1). The obtained \red {powder} X-ray diffraction (XRD) pattern of as-prepared $\rm K_2Ni_2TeO_6$ (Fig\red {ure S2} and Table \red {S}2) could be indexed to the hexagonal lattice adopting the P2-type framework. \cite{Masese2018} $\rm K_2Ni_2TeO_6$ displays a layered crystal structure comprising $\rm K$ atoms coordinated with oxygen atoms interposed between slabs of $\rm NiO_6$ and $\rm TeO_6$ as illustrated by Fig\red {ure} 1a. The $\rm NiO_6$ slab is composed of divalent nickel atoms \red {(as affirmed by X-ray absorption spectroscopy (Figure S3))} coordinated to six oxygen atoms, whereas the $\rm TeO_6$ contains $\rm Te$ atoms also coordinated to six oxygen atoms. As shown in Fig\red {ure} 1b, a honeycomb configuration of $\rm NiO_6$ and $\rm TeO_6$ octahedra is formed with each $\rm TeO_6$ surrounded by six $\rm NiO_\red {6}$ octahedra. In each unit cell, there are two repetitive layers of $\rm NiO_6$ and $\rm TeO_6$ octahedra that are separated by $\rm K$ atoms in a prismatic coordination with adjacent oxygen atoms. Even though humps indicating the presence of defects were detected (Fig\red {ure S}2), they could not be indexed using XRD, necessitating further characterisation using TEM.

\begin{figure*}
\centering
  \includegraphics[width=0.8\textwidth]{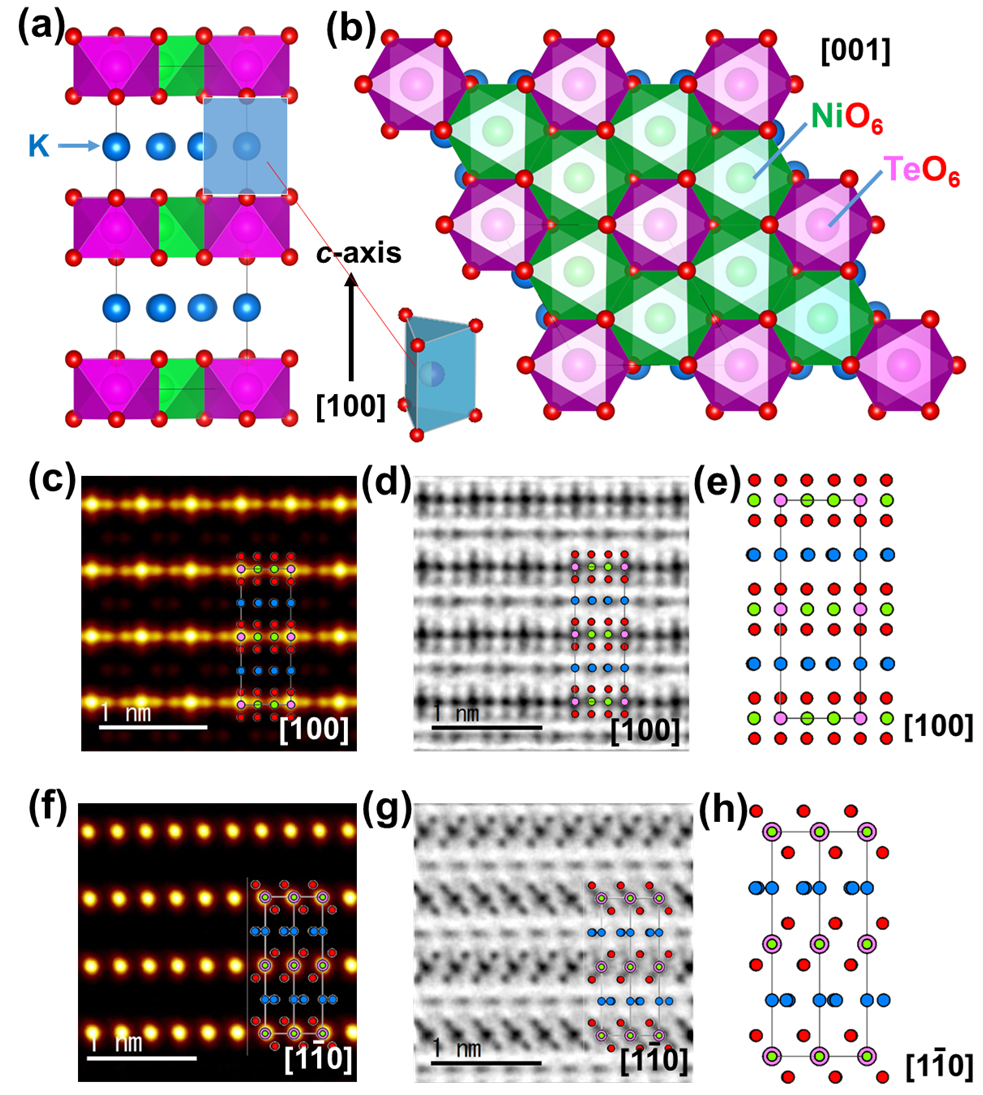}
  \caption{Visualisation of the honeycomb configuration of metal atoms in P2-type $\rm K_2Ni_2TeO_6$ along the $c$-axis. \red { (a)} Crystal structure of $\rm K_2Ni_2TeO_6$ along [100] projection showing the P2-type coordination (prismatic coordination of the $\rm K$ atoms and 2 honeycomb slab layers in the unit cell (shown in black). Throughout the figures, $\rm K$ atoms are shown in blue, $\rm Te$ in pink, $\rm Ni$ in green, and $\rm O$ in red. \red {(b)} Projection along [001] showing the honeycomb arrangement of $\rm Ni$ atoms around $\rm Te$ atoms. \red {(c)} High-angle annular dark-field scanning transmission electron microscopy (HAADF-STEM) image taken along [100] zone axis showing the ordering sequence of $\rm Ni$ and $\rm Te$ atoms corresponding to the P2-type stacking. Inset shows an atomistic view of the crystal structure, for clarity. \red {(d)} Annular bright-field (ABF)-STEM image along [100] zone axis showing also the arrangement of potassium atoms. \red {(e)} Rendering of the P2-type stacking of $\rm K_2Ni_2TeO_6$ along the [100] direction. \red {(f)} Visualisation (along the [1-10] zone axis) using HAADF-STEM, and \red {(g)} ABF-STEM. \red {(h)} Projection of the crystal structure along [1-10], affirming the polyhedral representation of P2-type stacking of atoms as shown in \red {(f)} and \red {(g)}.}
  \label{Fig_1}
\end{figure*}

Therefore, to attain more \red {nanostructural} information on the honeycomb ordering and slab stacking sequences, aberration-corrected scanning transmission electron microscopy (STEM) was employed on $\rm K_2Ni_2TeO_6$. Considering that prolonged electron beam irradiation gradually leads to sample degradation (Fig\red {ure S}4), the sample was exposed to electron beams for short periods, sufficient enough to obtain reliable high-resolution TEM (HRTEM) imaging without compromising the stability of the crystal structure (details in the \red {EXPERIMENTAL} section). For clarity, the contrast ($I$) of the STEM images are proportional to the atomic number ($Z$) of elements along the atomic arrangement (where $I \infty Z^{1.7} \approx Z^2$). \cite{pennycook1988, pennycook2006} The P2-type stacking of $\rm K_2Ni_2TeO_6$ is explicitly verified by the atomic-resolution high-angle annular dark-field scanning TEM (HAADF-STEM) images viewed along the [100] zone axis as illustrated by Fig\red {ure} 1c. The brighter and bigger yellow spots correspond to Te atoms ($Z = 52$), whilst the smaller yellow spots represent $\rm Ni$ atoms ($Z = 28$). As expected for a perfectly ordered P2-type honeycomb stacking sequence, the $\rm Te$–$\rm Ni$–$\rm Ni$–$\rm Te$ sequence along the $b$-axis is apparent, as illustrated by the crystal model (Fig\red {ure} 1e). The arrangement of the constituent atoms of $\rm K_2Ni_2TeO_6$ is further validated by atomic-resolution elemental mapping, taken along the [100] zone axis (Fig\red {ure S}5). Close inspection using annular bright-field (ABF)-STEM, as shown in Fig\red {ure} 1d, shows $\rm K$ atoms ($Z = 19$) positioned between the $\rm NiO_6$ and $\rm TeO_6$ slabs in a sandwich-like arrangement, as evinced by the atomistic model in Fig\red {ure} 1e. The $\rm K$ atoms are also noted to appear in varying contrasts (Fig\red {ure} 1d), indicating their occupation of crystallographically distinct sites with varying occupancies (\red {Table S}2), as can be further affirmed by HAADF- and ABF-STEM images taken along the [1-10] zone axis (Fig\red {ure}s 1f and 1g). A crystal model of the P2-type $\rm K_2Ni_2TeO_6$ viewed along the [1-10] zone axis is provided in Fig\red {ure} 1h. 

\begin{figure*}
\centering
  \includegraphics[width=\textwidth]{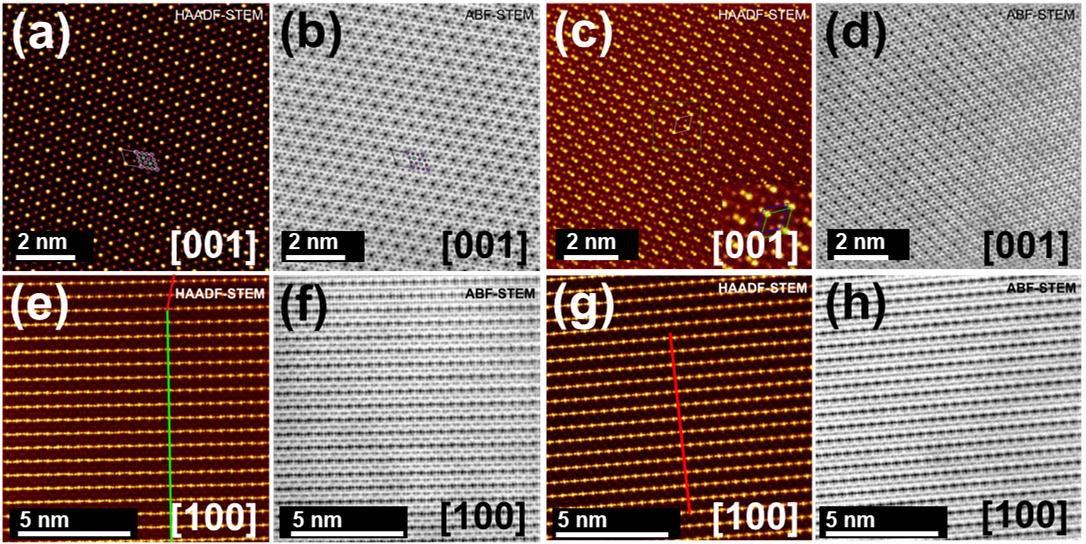}
  \caption{Visualisation of the honeycomb configuration of metal atoms in P2- and P3-type $\rm K_2Ni_2TeO_6$ along the $c$-axis and $a$-axis. \red {(a)} HAADF-STEM images along the [001] zone axis showing an ordered honeycomb arrangement of $\rm Ni$ (dark red spots) around the $\rm Te$ atoms (shown as bright yellow spots). Inset shows a polyhedral atomic view of the crystal structure, for clarity to the readers. \red {(b)} Corresponding ABF-STEM images along the [001] zone axis showing the atomic arrangement of $\rm Ni$ and $\rm Te$ atoms, in addition to atoms of lighter atomic mass such as potassium. \red {(c)} HAADF-STEM image taken along the [001] zone axis for a honeycomb slab part that shows doublets of both bright and dark spots (as shown further in inset), indicative of shearing (sliding) of the honeycomb slab (hence, the existence of stacking faults) along the $ab$ plane with a slab-to-slab translation vector of [1/3 1/3 0]. \red {(d)} Corresponding ABF-STEM micrographs taken along the [001] zone axis. \red {(e)} HAADF-STEM image taken along the [100] zone axis showing a new type of stacking (arrangement of which is shown in red line) that is distinct from the P2-type stacking (shown in green line). \red {(f)} Corresponding ABF-STEM images showing the arrangement of potassium atoms between the honeycomb slabs along the [100] zone axis. \red {(g)} HAADF-STEM micrographs of the domains adopting the new P3-type stacking along the [100] direction. The red line is a guide to the eye. \red {(h)} Corresponding ABF-STEM images taken along the [100] zone axis showing the arrangement of potassium atoms sandwiched between the P3-type stacking of honeycomb metal slabs.}
  \label{Fig_2}
\end{figure*}

A defect chemistry involving the cationic mixing of transition metal has been reported in honeycomb layered oxides such as $\rm Na_3Ni_2SbO_6$, where Ni and Sb have been observed to swap their crystallographic site positions.\cite{Xiao2020} To investigate this behaviour in $\rm K_2Ni_2TeO_6$, HAADF-STEM images of the cathode material viewed along the [001] zone axis were obtained as shown in Fig\red {ure} 2a. The corresponding ABF-STEM images (Fig\red {ure} 2b) explicitly confirm the honeycomb configuration of $\rm Ni$ atoms (depicted by the dark red spots) around $\rm Te$ atoms (brighter yellow spots) alongside $\rm K$ atoms overlapping with oxygen atoms ($Z = 8$). In this study, cationic mixing of $\rm Ni$ and $\rm Te$ was not detected in any of the crystallites investigated. The honeycomb arrangement of the $\rm Te$ and $\rm Ni$ atoms is further affirmed by atomic-resolution elemental mapping, taken along the [001] zone axis (Fig\red {ure S}6). Crystallite domains with doublets of $\rm Te$ and $\rm Ni$ spots appear to form a peculiar diagonal-like pattern amongst some particles of $\rm K_2Ni_2TeO_6$ as shown in Fig\red {ure}s 2c and 2d. The enlarged STEM image (Fig\red {ure} 2c) clearly shows that, in these domains, adjacent honeycomb slabs deviate from the perfect vertical alignment of $\rm Te$ and $\rm Ni$ atoms (Fig\red {ure}s 1e and 1h) along the [100] and [1-10] zone axes. Based on the slab stacking sequence along the $c$-axis ([001] direction), the adjacent slab shifts (by a translation vector of [1/3 1/3 0]) suggest that the new stacking variant in pristine $\rm K_2Ni_2TeO_6$ material may best be observed along the $a$-axis ([100]) or $b$-axis ([010]).

Therefore, to discern the new stacking variants, HAADF-STEM and ABF-STEM images of $\rm K_2Ni_2TeO_6$ were taken along the [100] zone axis, as seen in Fig\red {ure}s 2e and 2f. The initial P2-type slab stacking (shown in green line) can be observed as well as a new slab stacking domain that appears to shift along the $b$-axis. The new slab stackings are further elaborated by HAADF-STEM images shown in Fig\red {ure} 2g which show the $\rm Te$-atom arrangement to be repeatable after three $\rm K$ layers along the $c$-axis (inset). The arrangement of the new stacking domain \textit{vis-\`{a}-vis} the occupancy of the potassium atoms coordinated with oxygen atoms is also clearly seen in the corresponding ABF-STEM images (Fig\red {ure} 2h).

\begin{figure*}
\centering
  \includegraphics[width=0.7\textwidth]{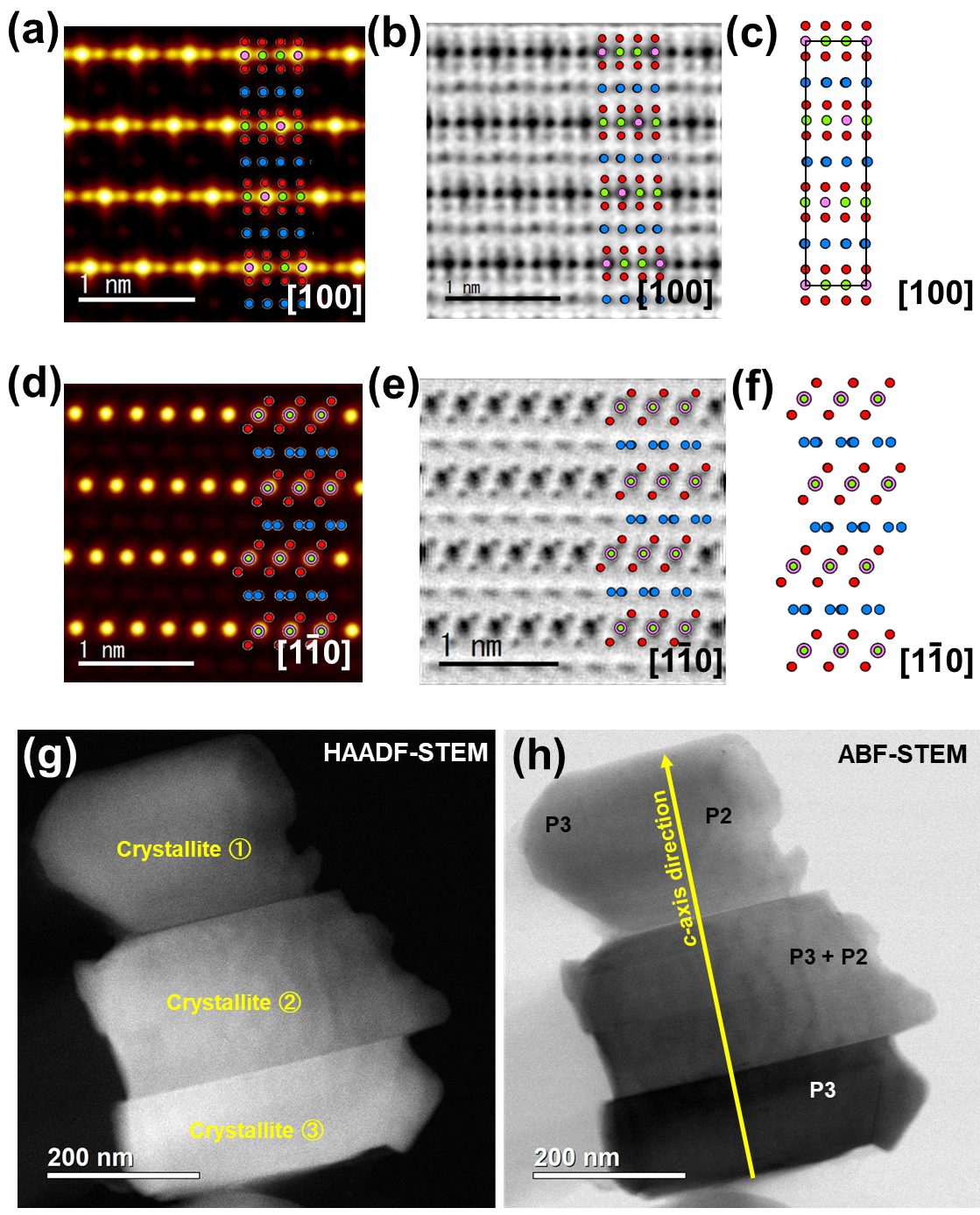}
  \caption{The newly identified P3-type stacking sequence in $\rm K_2Ni_2TeO_6$. \red {(a)} Typical HAADF-STEM image taken along [100] zone axis showing the ordering sequence of $\rm Ni$ and $\rm Te$ atoms corresponding to the P3-type stacking. Inset shows an atomistic depiction of the crystal structure, for the sake of clarity. \red {(b)} Corresponding ABF-STEM images taken along [100] zone axis showing also the arrangement of potassium atoms. \red {(c)} Representation of the crystal structural framework of P3-type $\rm K_2Ni_2TeO_6$ along the [100] direction. \red {(d)} Visualisation (along the [1-10] zone axis) using HAADF-STEM, and \red {(e)} ABF-STEM. \red {(f)} Projection of the crystal structure along [1-10], affirming the polyhedral representation of P3-type stacking of atoms as shown in \red {(d)} and \red {(e)}. \red {(g)} Low magnification HAADF-STEM micrographs of a section of crystallites aligned in various zone axes, as revealed by the varying degree of contrasts. \red {(h)} Corresponding ABF-STEM images showing crystallites with entirely P3-type stackings and some with a mixture of both P2-type and P3-type stackings. A vast majority of the crystallites adopted the P2-type stacking (major stacking phase). \red {The contrast difference between the upper, middle, and lower particles arises from the significant orientation changes upon rotation around the $c$-axis.} Details relating to the acquisition of the magnified images are furnished in the \red {EXPERIMENTAL} section.}
  \label{Fig_3}
\end{figure*}

To further assess the nature of the new stacking, atomic-resolution HAADF-STEM and ABF-STEM images were taken as shown in Fig\red {ure}s 3a and 3b. A new sequence with $\rm Te$–$\rm Ni$–$\rm Ni$–$\rm Te$ alternating along the $c$-axis is formed, indicating the formation of a new P3-type stacking by the slab-to-slab transition vector [1/3 1/3 0] relative to the main crystallographic axes, as displayed in the crystal model shown in Fig\red {ure} 3c. The new P3-stacking sequence was further verified by STEM images taken along the [1-10] zone axis (Fig\red {ure}s 3d and 3e) and a projected crystal model (Fig\red {ure} 3f). The occupancy of K atoms in the various crystallographic sites of the new P3 stacking is also noted to be different, as indicated by the varying contrasts of the $\rm K$ atoms coordinated with oxygen atoms along wavy-like columns in the $ab$ plane as illustrated by the crystal model in Fig\red {ure} 3c. This is further ascertained by the ABF-STEM images. By locating the position of oxygen atoms along [1-10] zone axis, the salient atomic structural differences between P2- and P3-type frameworks can be distinguished. The oxygen atoms are arranged diagonally in a zig-zag orientation along the $c$-axis in the P2-type framework (Fig\red {ure} 1g), whilst for the P3-type variant framework the oxygen atoms are aligned in the same direction on the $ab$ plane (Fig\red {ure} 3e). This observation was further corroborated by the low-magnification STEM images (Fig\red {ure}s 3g and 3h), where P3-type crystallites were located next to previously reported P2-type stackings. These observations not only demonstrate the emergence of a new type of stacking (P3-type) in pristine $\rm K_2Ni_2TeO_6$, but also explicitly show the existence of stacking variants (faults or disorders). Even so, an extensive examination into their stacking sequences is still necessary for a deeper insight into their crystallographic information.

As such, selected area electron diffraction (SAED) measurements were duly performed, exhibiting patterns of a crystallite with P2-stacking sequence along the [1-10] and [100] zone axes, as shown in Fig\red {ure}s. S7a and S7c. The corresponding simulations are shown in Fig\red {ure}s S7b and S7d, revealing a good match with the experimental diffractograms indexable to the hexagonal lattice of the P2-type phase. For comparison, the SAED patterns of a crystallite with P3-type stacking sequence are provided in Fig\red {ures} S7e and S7g along with their corresponding simulations (Fig\red {ures} S7f and S7h). The patterns of the domain with the P3-type stacking along [1-10] and [010] zone axes (Fig\red {ure}s S7e and S7g), reveal arrays of pseudo-hexagonal symmetry dots that could be indexed into a slightly-distorted hexagonal cell with the approximate lattice parameters: \red { $a \approx 5.26 \pm{0.05}$}  \AA, \red {$c \approx 18.70 \pm{0.02}$ \AA} (interslab distance of $6.23 \pm{0.06}$ \AA). The P3-type lattice parameters are very close to the lattice parameters obtained from the P2-type stacking: \red {$a \approx 5.25 \pm{0.05}$ \AA}, \red {$c \approx 12.44 \pm{0.1}$ \AA} (interslab distance of \red {$6.22 \pm{0.06}$ \AA}). Indeed, this explains why this new stacking (P3-type) variant was undetected by the bulk XRD analyses owing to its isostructurality and exacerbated further by its crystallite concentration which is presumably rather subtle to be sufficiently distinguished. In the crystallites analysed, P2-type domains were predominantly seen with P3-type domains seldom being detected as shown in Fig\red {ures} 3g and 3h. A further inspection of the electron diffractograms of P3-type stacking (shown in Fig\red {ure}s S7e and S7g) indicate angle deviations $77.7^\circ$ and $82^\circ$ from the adjacent main diffraction spots along the [1-10] and [010] zone axes, respectively. Typically, angle deviations are expected to fall around 90$^\circ$ with an accuracy range $ca$. 1\% as such errors arising from experimental flaws can be ruled out from the inclinations observed in the electron diffractograms. This indicates that the P3-type stacking lattice is slightly distorted.

Further investigations into the nature of the pristine $\rm K_2Ni_2TeO_6$ revealed for the first time, \red {nano}structural defects entailing the disappearance or bending of the honeycomb slabs and potassium layers. In the low-magnification STEM images of the new P3-type stacking domains (Figures 4a and 4b), various crystallites oriented in different zone axes seem to be separated by grain boundary lines. However, close scrutiny of the HAADF-STEM images (Fig\red {ure} 4c) reveal the warping or bending of the honeycomb slabs. Corresponding ABF-STEM images (Fig\red {ure} 4d) further show similar warping, that could be mistaken for twin or tilt boundaries within the potassium layers. Moreover, high-magnification STEM images (Fig\red {ure}s 4e and 4f) show an undulating topology of both the honeycomb slabs and potassium layers along the $b$-axis, indicating defects that relate to curvature of the honeycomb slab surface. Such defects are exceedingly rare in layered oxides, prompting deeper scrutiny into the $\rm K_2Ni_2TeO_6$ crystal structure.

\begin{figure*}
\centering
  \includegraphics[width=0.65\textwidth]{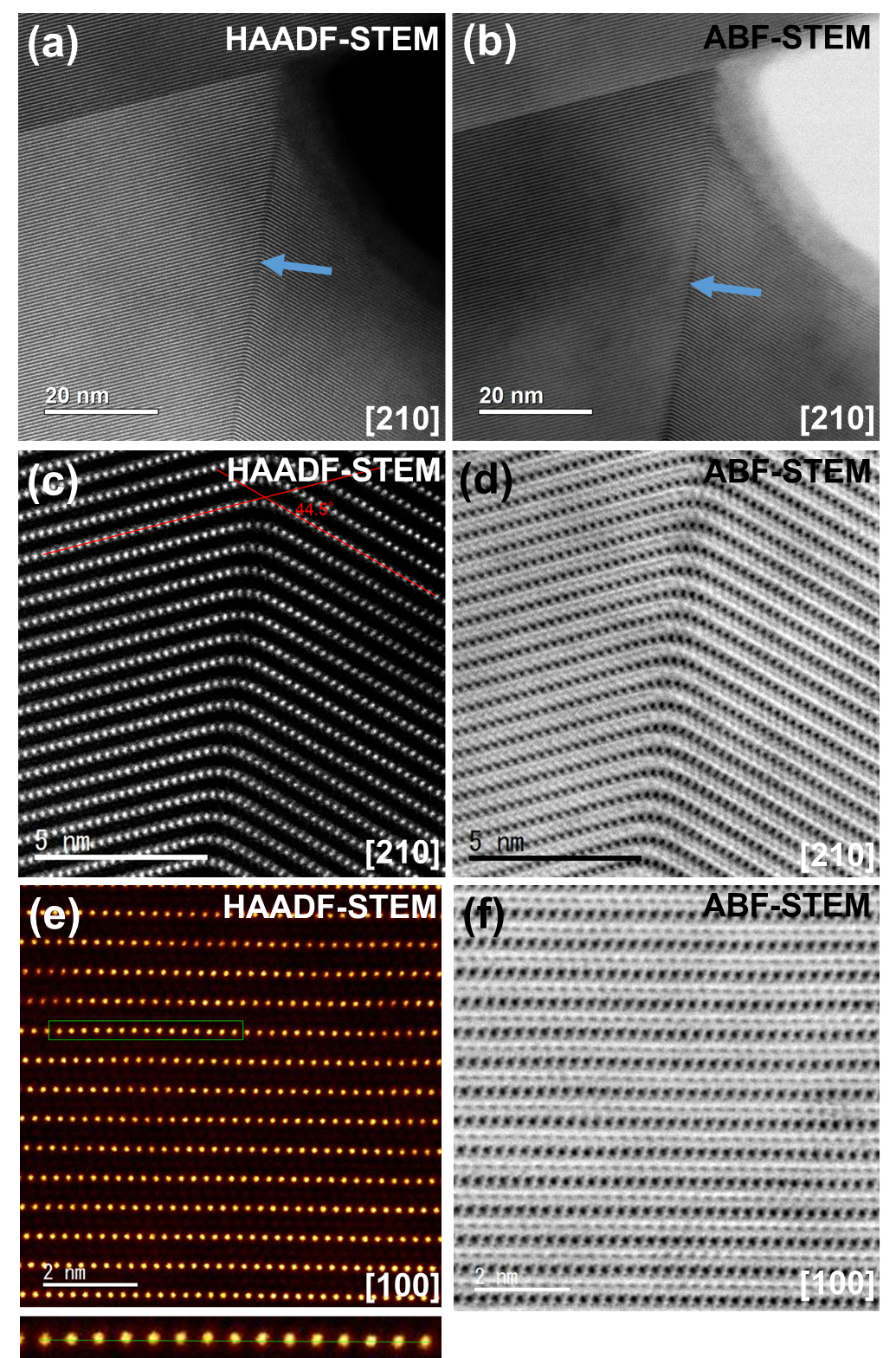}
  \caption{Topological defects related to strong curvature in P3-type $\rm K_2Ni_2TeO_6$. \red {(a, b)} High-resolution TEM micrographs of a section of P3-type stacking that shows curvature defects along [210] zone axis (underpinned using arrows). \red {For the sake of clarity, disparate contrasts emerge owing to the existence of a rotating grain boundary around the $c$-axis, in addition to the changing crystal orientation.} \red {(c)}  HAADF-STEM images along [210] zone axis showing the bending of the honeycomb slab layer at an angle of $44.5^\circ$ relative to the horizontal planes (as highlighted in red) and \red {(d)} Corresponding ABF-STEM images showing the bending of the alkali atoms layer (in this case, potassium). \red {(e)} HAADF-STEM images taken along [100] direction showing the undulating nature of the honeycomb metal stabs (as highlighted in inset (green line)) and \red {(f)} Corresponding ABF-STEM images also showing the undulating nature of the potassium atom layers across the honeycomb slab surface.}
  \label{Fig_4}
\end{figure*}

Topological defects such as (Taylor's) dislocations and disclinations, where translational and rotational symmetry of the atoms in the crystal are destroyed, have often been associated with beneficial physicochemical properties amongst oxides. \cite{Sandiumenge2019} As crystalline symmetries are intricately linked with momentum and angular momentum conservation law, such violations of can profoundly affect the dynamics of alkali cations during (de)intercalation processes.\cite{Kanyolo2020b} Generally, the origin (technically, dislocation core) of dislocations in the crystal lattice, is denoted by the symbol `T'. Figure 5a shows a HAADF-STEM image of a P2-type stacking domain along the [100] zone axis, where a curvature appears as a result of part of the honeycomb slab laying out of position (highlighted as `T'). The corresponding ABF-STEM images reveal considerable distortions on the $\rm K_2Ni_2TeO_6$ lattice as shown in Fig\red {ure} 5b. Edge dislocations on the lattice are clearly discerned as highlighted in Fig\red {ure}s 5c and 5d respectively. The images indicate that these dislocations are created by a shift along the $ab$-plane with a translation vector [1/3 1/3 0]. In principle, the magnitude and direction of the lattice distortion arising from such edge dislocation can be represented using a Burgers vector (b). The Burgers vector of the edge dislocation imaged in Fig\red {ure}s 5a and 5b was determined to be [0 1 0], spanning along the $b$-axis (slip plane). As for the edge dislocations appearing in Fig\red {ures} 5c and 5d, the Burgers vector was determined to be [1/3 1/3 1/2] spanning along the $b$-axis. Other unique edge dislocations detected during the analysis were determined to shift with a Burgers vector of [–1/3 –1/3 1/2] as indicated in Fig\red {ure} S8. It is noteworthy that, despite proving elusive, other shifts occurring along the $a$-axis cannot be ruled out. It must also be noted that the diversity and unpredictability of the edge dislocations observed in our investigation are unique to P3-type structures and are extremely rare amongst oxides.

\begin{figure*}
\centering
  \includegraphics[width=\textwidth]{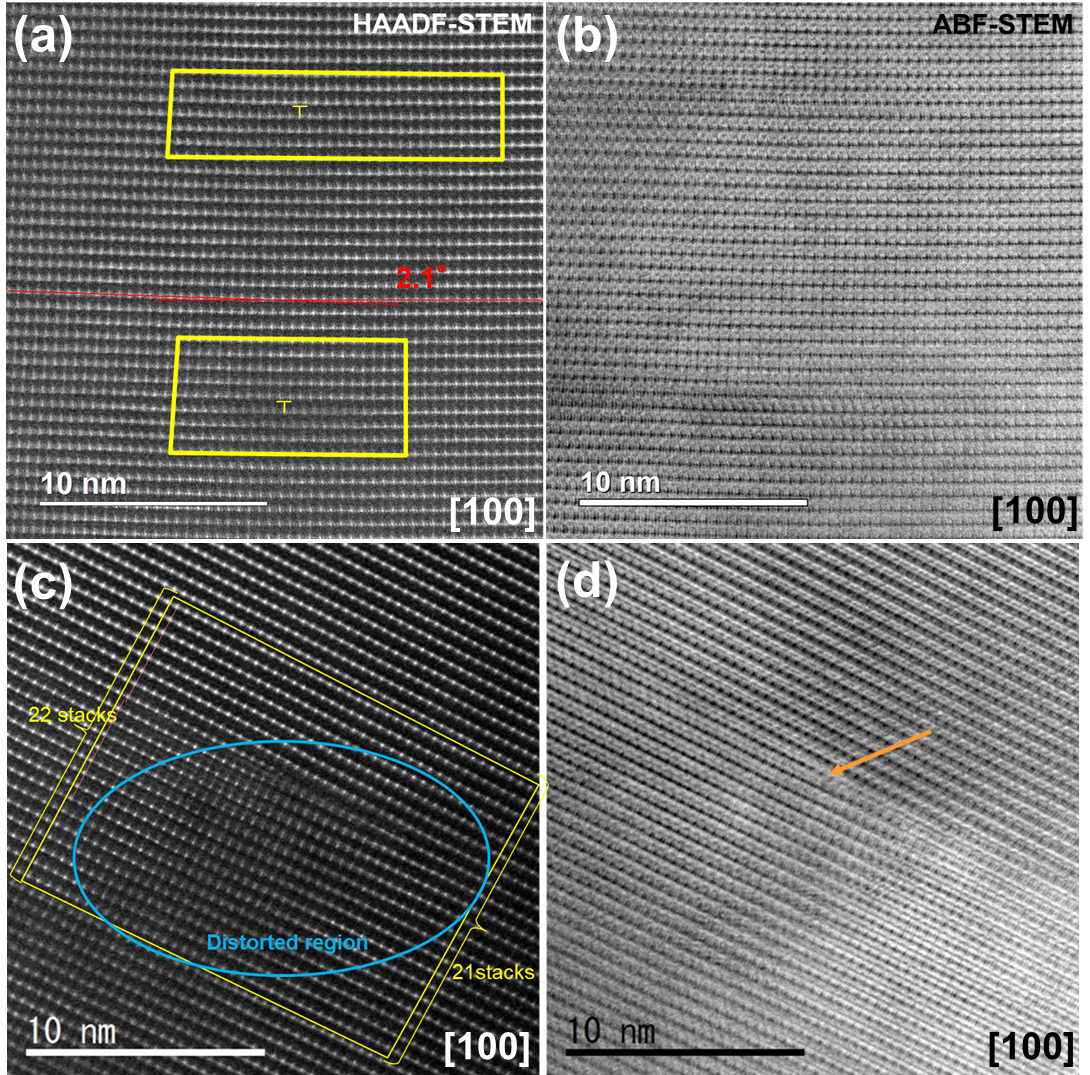}
  \caption{Topological defects related to weak curvature in P2-type $\rm K_2Ni_2TeO_6$. \red {(a)} HAADF-STEM images of P2-type stacking along the [100] zone axis, showing the bending of the honeycomb slabs owing to the formation of defects (in particular, edge dislocations). The point of origin of the dislocation is shown as `T'. For clarity, the `T' shape labels the slab mismatch direction in the edge dislocation core. \red {(b)} Corresponding ABF-STEM images along the [100] zone axis also showing the bending of the potassium atom layers. \red {(c)} HAADF-STEM images of a distorted honeycomb slab section of the P2-type stacking along the [100] zone axis and \red {(d)} the corresponding ABF-STEM images along the [100] zone axis highlighting the distorted slab where edge dislocations have occurred (highlighted by an arrow).}
  \label{Fig_5}
\end{figure*}

The edge dislocations (Fig\red {ure} 5) and curvature (Fig\red {ure} 4) identified, may provide insight into the $\rm K^+$ diffusion mechanisms of $\rm K_2Ni_2TeO_6$ particularly during (de-)intercalation processes. Generally, the mechanism generating correlations between stacking sequences and topological defects such as ion vacancies and atomic rearrangements is often analysed within the context of slab shearing/gliding/sliding, which need not include curvature deformations.\cite{Wang2018, blangero2005, zhang2019gliding, Ma2015} However, our work explicitly demonstrates that curvature is an integral part of such topological defects. In fact, in a previous work, it has been theorised that there is a direct relation between such curvatures and the vacancy of cations such as $\rm K^+$.\cite{Kanyolo2020b, Kanyolo2020a} Particularly, the integral of the 2D charge density of ionic vacancies ($\rho$) over a given patch of $area$ previously occupied by $\rm K^+$ ions ($\int \rho\, d(Area) = n$) is equal to the number of cationic vacancies ($n$) in the given $area$ patch. This is analogous to the well-known Gauss-Bonnet theorem ($\int K\, d(Area)$) in 2D geometry, that relates curvature to topological `holes' in the layer, where the (Gaussian) curvature ($K$) is proportional to the charge density of the ionic vacancies ($\rho = -K/4\pi$). \cite{Allendoerfer1943} For clarity to readers, a rendition representing such a mechanism is provided in Fig\red {ure} S9. The `holes' act as cationic vacancies forming whenever sufficient activation energy is supplied in the material to offset the binding energy of cations that form the stable P2-type lattice, leading to stacking sequence phase transformations as has been noted in $\rm Na_3Ni_2SbO_6$ during $\rm Na^+$ (de)insertion. \cite{Wang2018, Ma2015} Similar to the aforementioned shearing/gliding/sliding mechanisms, sufficient activation energies can be provided by raising temperatures and/or by supplying other forms of energy-momentum via electrochemical processes within the material. In the case of the $\rm K_2Ni_2TeO_6$ reported in this work, the temperatures (above $800^\circ$ C) applied during synthesis of these materials, was sufficient to offset the binding energies within the stable P2-type lattice. \cite{Masese2018} Therefore, the phase transition to another stable P-type sequence (P3) in pristine $\rm K_2Ni_2TeO_6$ (as further shown schematically in Fig\red {ure} S10) can possibly be traced to thermally-induced geometric shifts of adjacent honeycomb slabs (shear transformations) and the formation of cationic vacancies during heating, which are accompanied by the topological defects and weak curvatures given in Fig\red {ures} 4 and 5. \red{However, the presence or absence of potassium vacancies could not be ascertained owing to the dislocations being accompanied by inter-layer displacement over a wide area (see Figure S11). The lattice strain hampered the clear visualisation of potassium atoms around the dislocation core.} The large ionic radius of $\rm K^+$ that gives rise to weaker interlayer bonds, also seemingly favours the retention of such topological defects and weak curvature even after adiabatic cool-down, as summarised in Fig\red {ures} 4 and 5. \red {More information relating to the weak curvature identified in this study is given in Figure S12.} HAADF-STEM images along the [210] zone axis in Fig\red {ure} 5c shows that $\rm K_2Ni_2TeO_6$ is extremely malleable even under large stresses (for instance, elastic deformation) which result in even larger curvatures defects.  We expect future work to encompass the influence of the synthesis conditions (thermal treatment temperature, annealing duration, precursor types, \textit{etc}.) on the density (or volume fractions) of the \red {nanoscale} defects in such honeycomb layered oxides, as has been shown in layered oxides such as $\rm Na_{0.7}MnO_2$.\cite{eriksson2003}


\red{\section{Conclusion}}

In summary, we utilise the higher-order atomic-resolution in aberration-corrected scanning transmission electron microscopy (STEM), to explicitly reveal \red {nano}structural disorders (variants) of honeycomb layers along $c$-axis in pristine $\rm K_2Ni_2TeO_6$. We identify unique topological defects and curvatures due to partially missing honeycomb slabs of $\rm NiO_6$, $\rm TeO_6$ octahedra as well as $\rm K$ atoms. Moreover, a new stacking variant with P3-type stacking sequence is for the first time discovered to coexist with the predominant P2-type stacking of $\rm K_2Ni_2TeO_6$. In this study, it is postulated that the occurrence of the \red {nanoscale} topological defects and curvature is correlated with missing atoms (dislocations) leading to the emergence of the P3-type stacking. We expect this work will prompt rekindled interest in the study of the defects, curvature and their role in the functionality of such honeycomb layered oxides \red {in applications related, for instance, to rechargeable batteries}.

\section*{Author contributions}

T. M. and H. S. contributed to the syntheses of the materials. Y.M., T.S., M.I. and T.T. performed the morphological characterisation of the materials using TEM. Y.M., G.K., T.M, H.S. and T.S. wrote the manuscript. G.K. and T.M. designed the graphic illustrations outlined in the work. G.K., H.S., T.M. and Y.M. provided input with the data analyses, helped with the discussion and assisted with the manuscript correction. T.T., M. I. and T.S. made the initiative to undertake this work. All authors have given approval to the final version of the manuscript. 

\section*{Data availability}

The data that support the findings of this study are available on request from the corresponding authors [T.M.], [Y. M.], [G. K.], [H. S.] and [T. S.].

\section*{Corresponding authors}

Correspondence to Titus Masese or Yoshinobu Miyazaki or Godwill Kanyolo or Hiroshi Senoh or Tomohiro Saito.

\section{Competing interests}

The authors declare no competing interests.

\begin{acknowledgement}
T.M. thanks Ms. Shinobu Wada and Mr. Hiroshi Kimura for the unrelenting support in undertaking the entire study. T. M. also gratefully acknowledges Ms. Kumi Shiokawa, Mr. Masahiro Hirata and Ms. Machiko Kakiuchi for their advice and technical help as we conducted the syntheses, electrochemical and XRD measurements. Dr. Minami Kato is thanked for the assistance in designing the graphical images. This work was conducted in part under the auspices of the Japan Society for the Promotion of Science (JSPS KAKENHI Grant Number 19K15685), Sumika Chemical Analyses Services (SCAS) Co. Ltd., National Institute of Advanced Industrial Science and Technology (AIST) and Japan Prize Foundation.
\end{acknowledgement}

\begin{suppinfo}

The following files are available free of charge at https://pubs.acs.org/doi/10.1021/acsanm.0c02601.
\begin{itemize}
  \item Supporting Information: Experimental details and figures showing the material characterisation \red {(XRD, SEM, TEM, ICP-AES and X-ray absorption spectroscopy (XAS) data)} of pristine $\rm K_2Ni_2TeO_6$. 
\end{itemize}

\end{suppinfo}

\bibliography{achemso.bib}

\end{document}